\documentclass[a4paper,11pt]{article}
\usepackage{jinstpub} 
\usepackage{siunitx}
\usepackage{lineno}

\title{\boldmath Application of the VMM3a/SRS: \\ A t0-less TWIN GEM-based TPC}

\author[c,m,*]{K.J. Flöthner}
\author[a,*]{F. Garcia}
\author[b,e]{B. Oberhauser}
\author[c]{F. Brunbauer}
\author[b]{M. W. Heiss}
\author[c,f,g]{D. Janssens}
\author[d]{B. Ketzer}
\author[c]{M. Lisowska}
\author[c,l]{, M. Meurer}
\author[c,d]{H. Muller}
\author[c]{E. Oliveri}
\author[c,h]{G. Orlandini}
\author[c]{D. Pfeiffer}
\author[c]{L. Ropelewsk}
\author[c,i]{J. Samarati}
\author[c]{F. Sauli}
\author[c,d]{L. Scharenberg}
\author[c]{M. van Stenis}
\author[c,k]{A. Utrobicic}
\author[c]{R. Veenhof}

\affiliation[a]{Helsinki Institute of Physics, University of Helsinki,Gustaf Hällströmin katu 2a,Helsinki,FI-00014,Finland}
\affiliation[b]{Paul Scherrer Institut,Forschungsstrasse 111, 5232 Villigen,Switzerland}
\affiliation[c]{European Organization for Nuclear Research (CERN),1211 Geneva 23,Switzerland}
\affiliation[d]{Physikalisches Institut, University of Bonn,Nußallee 12, 53115 Bonn,Germany}
\affiliation[e]{ETH Zürich, Swiss Federal Institute For Technology Institute for Particle Physics and Astrophysics (IPA),Otto-Stern-Weg 5, 8093 Zürich,Switzerland}
\affiliation[f]{Inter-University Institute for High Energies (IIHE), Belgium}
\affiliation[g]{Vrije Universiteit Brussel, 1050 Brussels, Belgium}
\affiliation[h]{Friedrich-Alexander-Universit¨at Erlangen-N¨urnberg, Schloßplatz 4, 91054, Erlangen, Germany}
\affiliation[i]{European Spallation Source ERIC (ESS), Box 176, SE-221 00 Lund, Sweden}
\affiliation[k]{Ruder Boˇskovi´c Institute Bijeniˇcka c. 54, 10000 Zagreb, Croatia}
\affiliation[l]{Ludwig Maximilian University of Munich, Am Coulombwall 1, 85748, Garching, Germany}
\affiliation[m]{Helmholtz-Institut f\"{u}r Strahlen- und Kernphysik, University of Bonn, Nu\ss{}allee 14-16, 53115 Bonn, Germany}

\emailAdd{karl.jonathan.floethner@cern.ch}
\emailAdd{Francisco.Garcia@helsinki.fi}

\abstract{
Integrating the ATLAS/BNL VMM3a ASIC (Application Specific Integrated Circuit) into the RD51/SRS (Scalable Readout System) provides a self-triggered continuous readout system for various gaseous detectors. 
Since the system allows flexible parameters, such as switching the polarity, adjusting electronics gain or different peaking times, the settings can be adjusted for a wide range of detectors.
The system allows particles to be recorded with a MHz interaction rate in energy, space, and time.

The system will be introduced in the beginning, and short examples will be given for different applications.
After, the Twin GEM TPC will be discussed in more detail to show the benefits of such a trigger-less system in combination with the Twin configuration.
Last, a few results for the tracking performance and the possibility to operate as a tracking telescope will be shown.
Thus, this presents the possibility of an extremely low material budget tracking system suitable for tracking from high to low-energy particle beams.
}


\arxivnumber{xxxxxxxxx} 

\begin{document}
\maketitle
\flushbottom

\section{Introduction}
\label{sec:intro}
The VMM3a \cite{Iakovidis:2693463} has been integrated into the SRS \cite{LUPBERGER201891} within the framework of the RD51 collaboration.
Since then, hardware, firmware and software have been improved, resulting in a flexible readout system \cite{VMM3aSRSdoc}, suitable for multiple applications.
It allows single-detector laboratory setups to study detector physics, debug the system, and optimize the noise.
An example would be fast X-ray imaging with a 2MHz interaction by utilizing the high rate capability of the system \cite{SCHARENBERG2021165576, PFEIFFER2022166548}.
Furthermore, the scalability of the systems allows an extension towards multiple detectors, building beam telescopes.
The RD51 VMM3a/SRS telescope, which consists of three compass-like~\cite{ALTUNBAS2002177} triple GEM tracking detectors, e.g., can provide track time resolutions below 2ns with track resolutions below 50um  \cite{Scharenberg:2860765} and can be extended to a 20-40m lever arm to optimize angular resolution.
Due to the rate capability, statistics of \num{e5} tracked Muons within one SPS spill of 5s is possible, minimising the time required for characterising a DUT.
High-rate studies can be performed up to 1MHz of pion rate using a focussed Pion beam.
This work will focus on showing one of the first operations of the system with the TWIN GEM TPC (HGB4-2), opening the application for ultra-low material budget tracking.
Hence, we will take this as a prime example and describe the operation of a TPC in twin configuration without external to, benefiting of the combination of the trigger-less readout with the Twin configuration.
Furthermore, the first results of the spatial resolution of the TWIN GEM TPC will be presented, by using the mean cluster information of both TPCs.
Finally, the 3D-tracking will be shown, opening the possibility of operating the setup as an ultra-low material budget tracking telescope.
All presented data has been taken within the RD51 testbeam campaign in 2023 at the H4 beamline\cite{H4Beamline} of the SPS at CERN.
A wide muon beam at 150GeV/c momentum was used.
The VMM3a Slow Control and Calibration Software~\cite{VMMsc} was used to control the electronics, tcpdump~\cite{tcpDump} was used for the data taking, and vmm-sdat~\cite{vmmsdat} was used for cluster reconstruction.


\section{Operation of t0-less TPC}
The concept of TWIN TPCs has been introduced for the Super-FRS tracking of heavy ions \cite{Sitar2015} to increase the rate capability of the conventional tracking TPCs and then extended for the high-rate operation to the GEM-based TPC~\cite{Garcia2015, Garcia2018}.

\begin{figure}[h!]
\centering 
\includegraphics[width=0.55\textwidth]{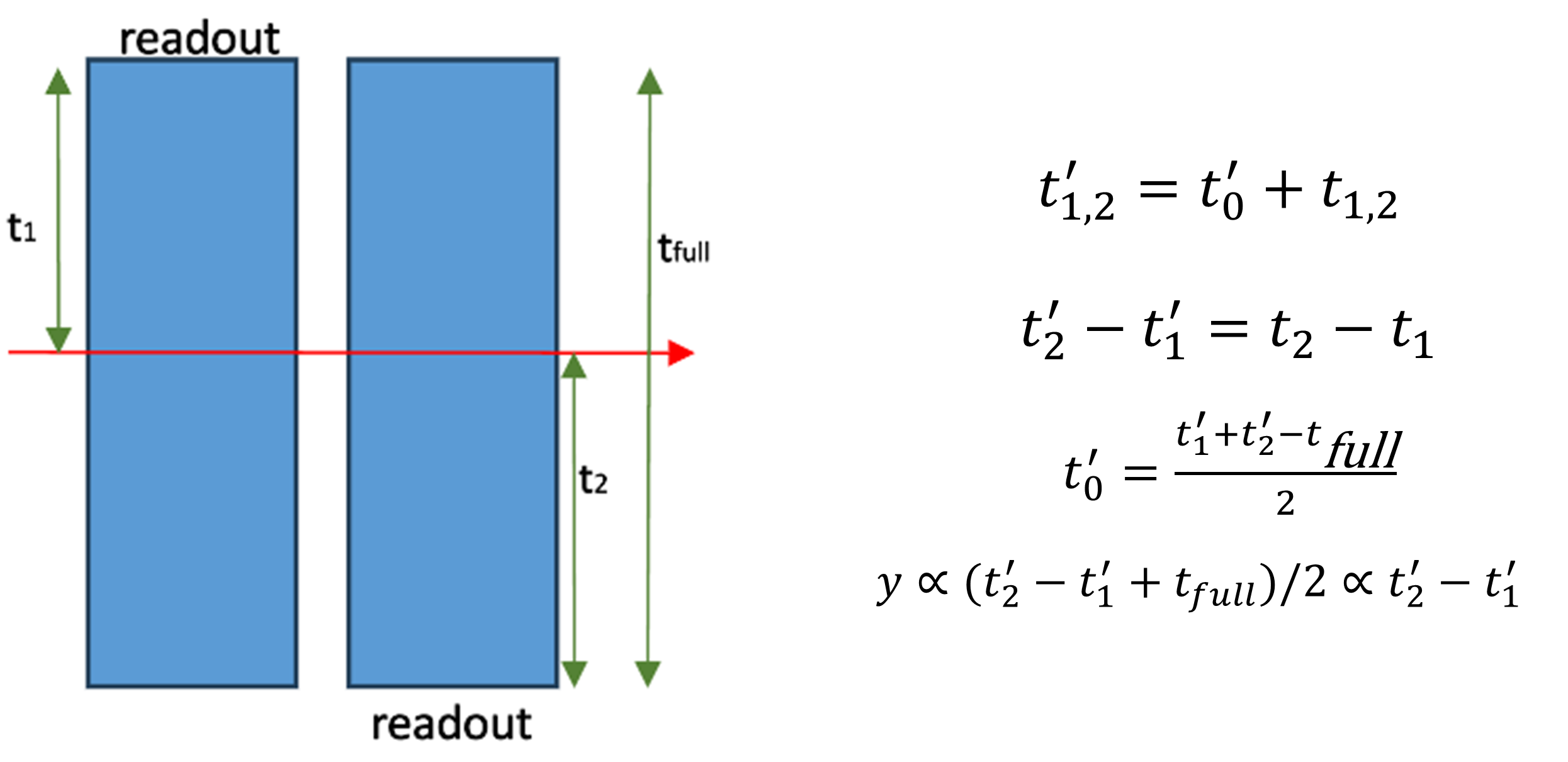}
\caption{\label{fig:t0schememath} Schematic of Twin configuration and resulting dependencies for drift times and y position.}
\end{figure}

This chamber has two GEM-TPCs inside a single vessel sharing the same gas volume, with one of them rotated $\SI{180}{\degree}$ in the middle plane with respect to the other, as shown in Fig.~\ref{fig:t0schememath}.

\begin{figure}[h!]
\centering 
\includegraphics[width=0.7\textwidth]{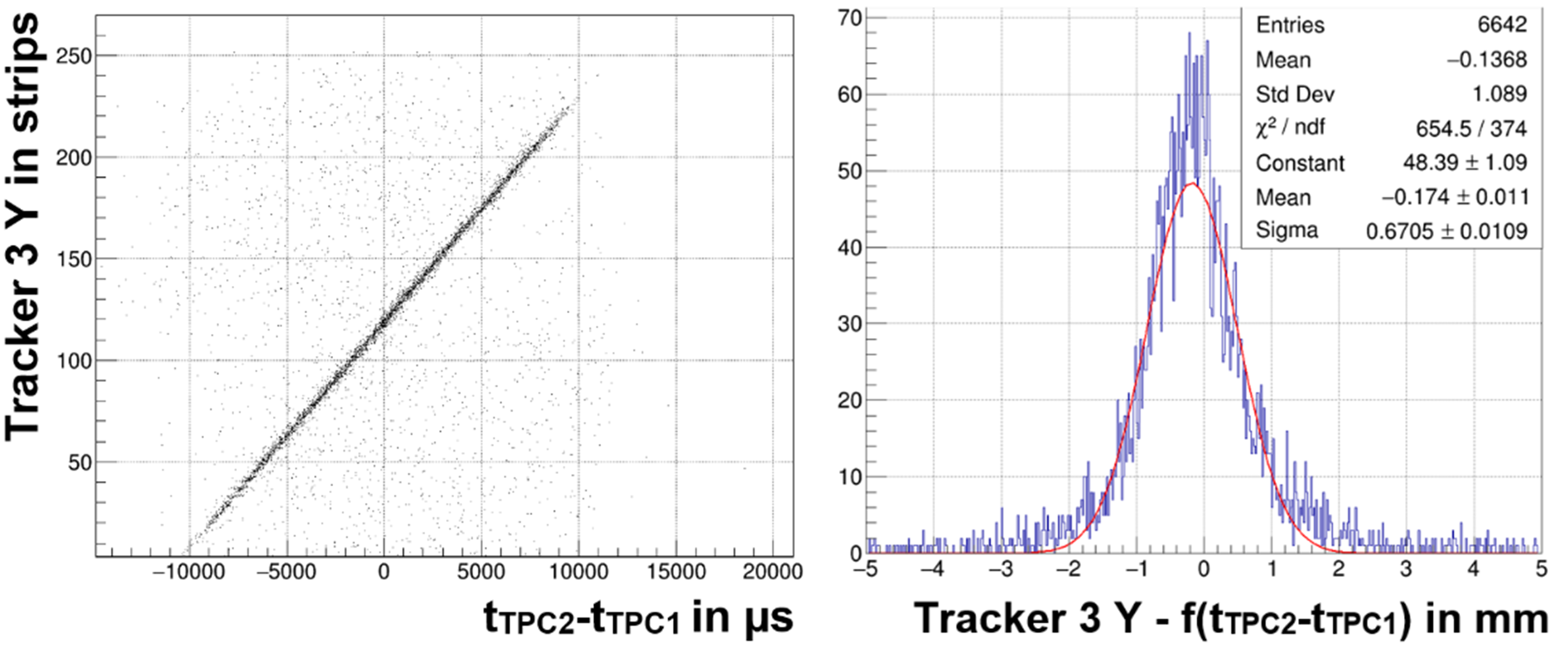}
\caption{\label{fig:Removet0} Left: Correlation between Y position of third tracking detector and time difference. Right: Residual distribution of position, reconstructed through a linear fit of correlation.}
\end{figure}

To calculate the position from the relative times $t_{1,2}$, the absolute $t'_{0}$ and drift time need to be known to calculate a Y position for each TPC.
However, one can remove the $t'_{0}$ dependence substracting the two absolute drift times $t'_{1,2}$ from each other, due to the geometrical dependence.
The absolute $t'_{0}$ can even be expressed by the absolute times $t'_{1,2}$ and the full drift time $t_\text{full}$.
Due to the self-triggered operation of the VMM3a/SRS, the absolute time of the DAQ is known for each strip of every event.
Therefore, after matching events between the TPCs, the position in Y is directly proportional to the difference in the absolute times of the TPCs.
Events are found by matching within a time window corresponding to a $\pm t_\text{full}$ window.
With the calculated t0, one can calculate the sum of both drift times (control sum), which has to correspond to the full drift time in case of a correct matching between the TPCs.
Hence, the control sum can be used to prove correct matching and can be used to resolve ambiguities for multiple tracks in $\pm t_\text{full}$.
The position dependence can be seen by plotting the time difference against the Y coordinate of one of the telescope trackers, e.g., T3, the third one.
The result is shown in Fig.~\ref{fig:Removet0} and gives a clear, linear correlation.
Through a linear fit, the time difference can be translated into a distance to calculate the residual distribution, as shown in Fig. \ref{fig:Removet0}.
This proves the t0-less operation, and given the low divergence of the team, an upper limit of the spatial resolution can be estimated at 671(11)um.
A more accurate calculation of the resolution will be shown in the following.

\section{Spatial resolution by mean cluster information}

The main observable under investigation is the spatial resolution of the HGB4-2, GEM-TPC in TWIN configuration with Ar/CO\textsubscript{2} (70/30 \%).
The reference tracks have been reconstructed from the telescope's three triple GEM tracking detectors as a reference.
Using a wide muon beam, the results will be a larger area. Therefore, the resolution depends on the drift velocity, the electronics' time resolution, and the drift field homogeneity of the active region.
The residual distributions can be seen in Fig.~\ref{fig:residualsPlane} where the position in X is calculated by the mean of the COG of the strip readouts of both TPCs and the Y position through the time difference of the meantime of clusters between the TPCs.
\begin{figure}[h!]
\centering 
\includegraphics[width=0.74\textwidth]{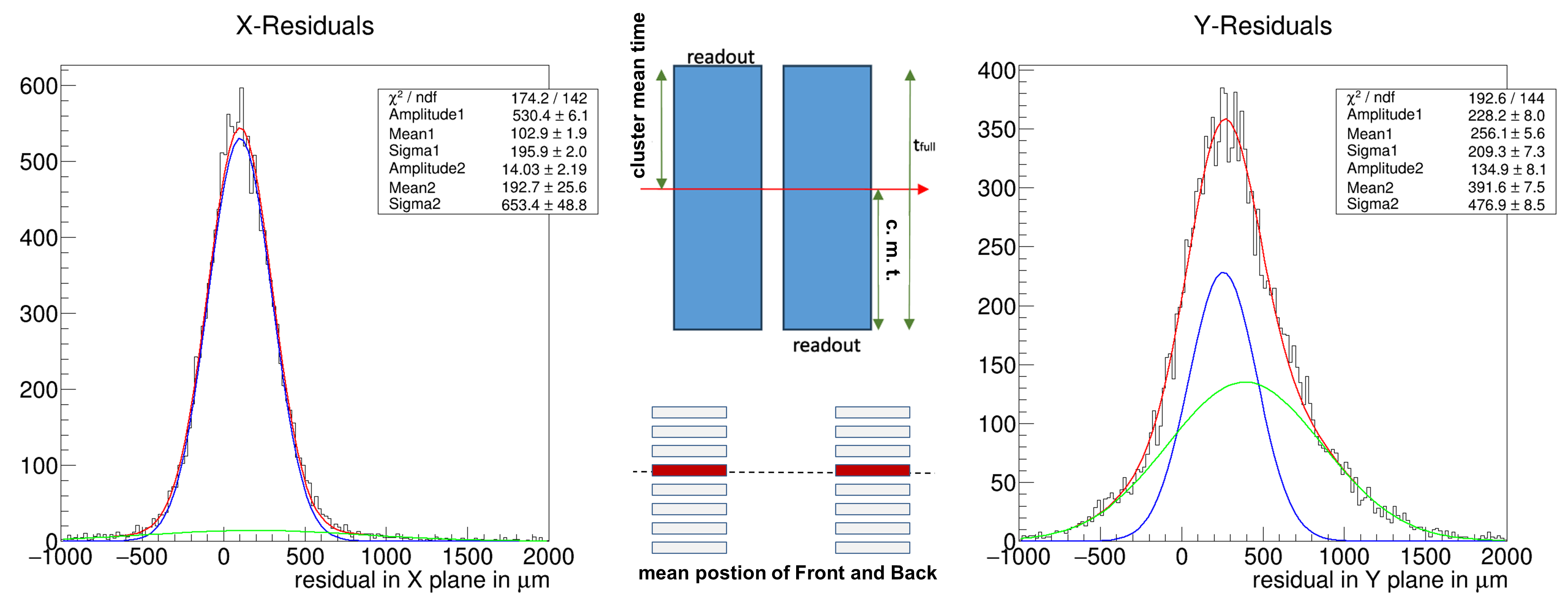}
\caption{\label{fig:residualsPlane} Left: Residual distribution in Y. Center: Schematics for position reconstruction. Right: Residual distribution in Y}
\end{figure}
The spatial resolution is estimated through a double Gaussian fit.
Using the combined sigma, the resolution in X is found to be 220.0(54)um and in Y to be 334.7(70).
The reasons behind the asymmetric distribution in the Y residual are unclear but might correlate to the drift field's uniformities.
For a better analysis, a detailed field map could help to apply a correction to mitigate such effects and improve the resolution.

\section{3D-tracking}
An optimal 3D-tracking would be performed with a pixelated readout to avoid a limitation of the dead time in individual channels ($\sim$200ns for an individual VMM3a channel).
The setup was tilted horizontally and vertically to overcome this with the 1D strips, introducing a time dependency of the hits in the z-position.
Previously, it was shown that the position of a traversing particle could be derived from the Twin TPC.
Being a TPC, extracting the angles of individual events and getting a full-track reconstruction should also be possible.
The reconstruction of the horizontal angle is straightforward by using the distance in strips between the two TPCs.
To correct for an offset between the TPCs, the distance between them in the nominal position must be considered.
The resulting horizontal angle for a strip pitch of 400um was calculated to be 4.73(15) degrees.

The angle reconstruction in Y is based on plotting and fitting the Y position of the hits within a cluster against the Z position of the hits within one cluster.
From this graph, three values can be extracted by selecting the fit range to cover only one of the TPCs or the full range.
An example event and the angle distributions of all tracks are shown in Fig. \ref{fig:residualsPlane}.
From this, the angle derived from the front TPC is 3.0(9) degrees, from the back TPC 3.3(9) degrees and from the whole system 3.5(3) degrees.
The distributions have roughly a factor of ten greater sigmas than the sigma reconstructed from the tracking of the telescope with 0.0375(6) degrees.
\begin{figure}[h!]
\centering 
\includegraphics[width=.99\textwidth]{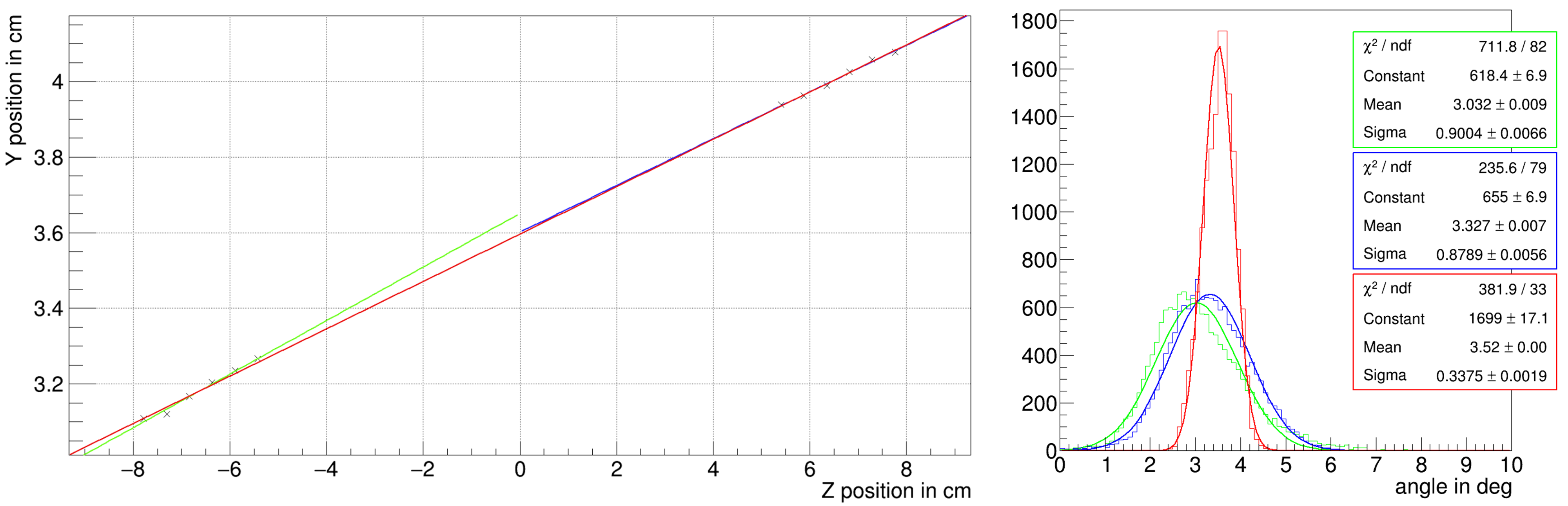}
\caption{\label{fig:residualsPlane} Left: Example for the reconstruction of a single event. Right: Distribution of vertical angles.}
\end{figure}
With this, it is proven that a full track reconstruction is possible with the TWIN GEM TPC (HGB4-2).
To improve the performance, further measurements with pixelated readout are planned.
\acknowledgments

We thank the RD51 test beam workgroup for facilitating the necessary infrastructure to carry out these measurements. 
Many thanks at GSI to C. Kaya, J. Kunkel, B. Voss, and H. Risch for the design of the detector, for providing the field cages, and to Christian Schmidt for facilitating the assembly infrastructure.
R. Turpeinen produced the flanges and window frames at the Helsinki Institute of Physics and the University of Helsinki mechanical workshop. J. Heino for helping with the quality assurance and framing.
Thanks to S. Rinta-Antila at the University of Jyväskylä's mechanical workshop for providing the housing of the detector.
We thank R. de Oliveira and the CERN microelectronics workshop for supporting and producing the GEM foils and frames.

\noindent
This work is sponsored by the Wolfgang Gentner Programme of the German Federal Ministry of Education and Research (grant no. 13E18CHA).

\noindent
This work has been supported by ETH Zurich and SNSF Grant No. 169133, 186181, 186158, 197346, 219485 (Switzerland).





\newpage
 \bibliographystyle{JHEP}
 \bibliography{biblio}
%





\end{document}